\documentclass{article}
\usepackage{emulateapj}
\usepackage{psfig}

\begin{document}

\title{Simultaneous {\it Chandra\/} and {\it RXTE\/} observations of the 
nearby bright Seyfert 2 galaxy NGC~4945}

\author{Chris Done}
\affil{Department of Physics, University of Durham, South Rd, DH1 3LE Durham,
UK}
\author{Greg M.~Madejski}
\affil{Stanford Linear Accelarator Center, 2575 Sand Hill Rd, Menlo Park,
CA 94025, USA}
\author{Piotr T.~\.{Z}ycki}
\affil{Nicolaus Copernicus Astronomical Center, Bartycka 18, 00-716 Warsaw,
Poland}
\author{Lincoln J. Greenhill}
\affil{Harvard-Smithsonian Center for Astrophysics, 60 Garden St, Cambridge,
MA 02138, USA}

\begin{abstract}

We analyze recent simultaneous {\it Chandra/RXTE\/} observations of
the Seyfert 2 galaxy NGC 4945. The unprecedented spatial resolution of
Chandra means we are able to separate the spectra of the nucleus,
starburst and superwind regions, while the RXTE data extend the
spectrum to higher energies. The extreme absorbing column of $N_H\sim
4\times 10^{24}$ cm$^{-2}$ means that the nucleus is only seen
directly above 8--10 keV, while the lower energy spectrum from the
nuclear region in {\it Chandra} is dominated by reflection. By
contrast, the superwind is dominated by emission from hot plasma, but
the starburst region contains both hot plasma and reflection
signatures.  To form a reflected spectrum requires that the starburst
region contains clumps of cool, optically thick material, perhaps star
forming cores, which are irradiated by 7--10 keV photons from the
nucleus. Since photons of this energy are obscured along the line
sight then this confirms the result of Madejski et al. (2000) that the
extreme absorbtion material is disk-like rather than a torus. However, the
IR/optical limits on the lack of high excitation emission lines show
that by contrast the lower energy photons from the nucleus are obscured
in all directions.  We discuss the complex absorption structure
revealed by these observations, and propose an an overall source
geometry in which the nucleus is completely embedded in material with
$N_H\sim 10^{23}$ cm$^{-2}$

\end{abstract}

\section{Introduction}

Our best current picture of Seyfert 2 galaxies relies on the Unified
Scheme, where all the main ingredients of the nucleus -- black hole,
accretion disc and broad line region -- are identical in all active
galaxies, but the classification depends on the orientation with
respect to the line of sight.  There is a small scale, 
geometrically thin accretion disk around the black hole,
while at distances larger than $\sim$ 1~pc there is a
geometrically thick molecular torus. The material in the disk and
torus provides intrinsic obscuration, such that for the viewing
directions intersecting the disk and/or torus, the intrinsic emission
of the AGN is modified by the absorbing material.  In particular, the
light from the broad optical / UV emission lines is obscured by dust,
while photoelectric absorption by gas and dust gives a low energy
cutoff in the soft X--ray emission. The magnitude of the absorption
depends on the column density of material in the line of sight. For
columns of $N_H > 10^{25}$ cm$^{-2}$ the central engine can be completely
obscured, though some small fraction of the intrinsic nuclear light
can be seen as a result of electron scattering by low density, ionized
gas; this gas would be located around the axis of symmetry of the
system, both within the torus and in a form of a wind emanating from
the nucleus.  

Such a picture is broadly consistent with a range of observed optical,
X--ray, radio and polarization properties of Seyfert 1 and 2 galaxies
(see e.g. the review by Veron-Cetty \& Veron 2000), but these data are
generally indirect as they do not resolve any of the structures
proposed.  One of the nearest AGN of any kind -- and thus the most
appropriate for a detailed study of the spatial structure to test
these Unified models -- is the nearby (3.7 Mpc; Mauersberger et
al. 1996) Seyfert~2 galaxy NGC~4945.  Spatial studies are further
aided by the strong absorption, with an equivalent hydrogen column
density of $\sim 4 \times 10^{24}$ cm$^{-2}$ (Iwasawa et al. 1993;
Done, Madejski, \& Smith 1996; Madejski et al. 2000), corresponding to
$\tau_{\rm Thomson}$ of $\sim 2.4$.  With this column, the nuclear
X-ray flux at energies corresponding to the Fe L and K$\alpha$ lines
is entirely absorbed, so the measured line fluxes originate in the
scatterer or in the obscuring medium, yet above $\sim 10$ keV, the
nuclear power law dominates the spectrum.  In fact, it is the
brightest Seyfert~2 galaxy above 20 keV, as measured by OSSE (Done,
Madejski, \& Smith 1996), RXTE (Madejski et al.\ 2000), and BeppoSAX
(Guainazzi et al.\ 2000).  The hard X-ray emission is rapidly
variable, on time scales of $\sim$ a day or less, implying that the
Thomson-thick absorption is probably confined to a structure which is
geometrically rather thin, i.e. more probably associated with the disk
rather than the torus (Madejski et al. 2000).  Importantly, NGC~4945
is an H$_2$O megamaser source (Dos Santos \& Lepine 1979), 
which traces underlying cool, dense molecular structures probably
within $\sim 1$~pc of the central engine. Unlike NGC\,4258, this 
emission does not give the smooth rotation curve expected from 
a well ordered disk, but assuming this irregular, 
clumpy distribution still traces orbital motion gives an estimate of
the central mass of $\sim 1.4 \times 10^6$ M$_\odot$
(Greenhill, Moran, \& Herrnstein
1997).  Knowing the intrinsic X--ray luminosity {\sl and} the mass of
the central source allows an estimate of the accretion rate in
Eddington units of 10\% (Greenhill et al.\ 1997; Madejski et al.\
2000), and aids in detailed dynamical studies of the source.

The ionized material filling the opening of the torus also emits 
recombination line and continuum radiation, generally resulting 
in strong Fe L and K$\alpha$ lines (Krolik \& Kallman 1987;
Band et al. 1990).  However, despite this theoretical work 
there is comparatively little known about the scattering region 
in Seyfert~2s.  This is mainly due to spatial confusion, as Seyfert 
nuclei often co--exist with nuclear starbursts/superwind 
activity, and these contribute to the soft X--ray emission and Fe L 
lines.  This is certainly the case for NGC~4945 (Heckman, 
Armus, \& Miley 1990;  Nakai 1989).  To disentangle the effects 
of the scattering region from the starburst and superwind, we need 
high spatial resolution, provided by the superior angular resolution 
available with {\it Chandra\/}.  

In this Paper, we report on the {\it Chandra\/} imaging observations of the
nuclear region of NGC~4945. These have already been published by
Schurch, Roberts \& Warwick (2002), but here we do a much more
detailed spatially resolved analysis. We also include simultaneous data
from RXTE in order to constrain the direct nuclear spectrum, and
present a plausible interpretation regarding the geometry of
the source which fits with both the IR/optical and X-ray constraints. 

\section{Observations}
\label{sec:obs}

\subsection{Chandra}

\begin{figure} 
\plotone{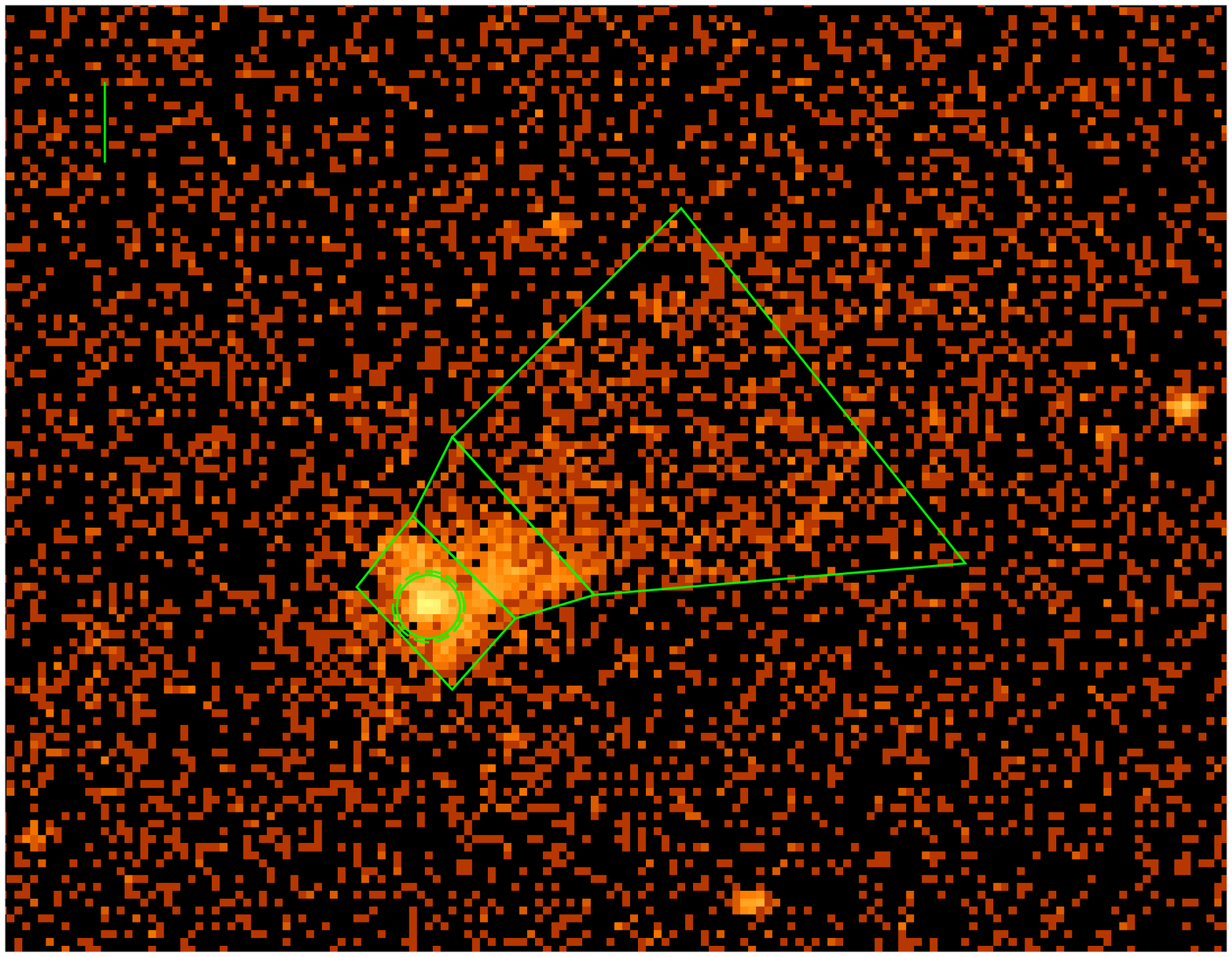}
\caption{Chandra image of NGC~4945, with the spatial regions marked
where the data for spectral analysis was extracted. 
North is up and East is to the left. The bar in the top left hand
corner is 5 arcsec long (corresponding to 90 pc at 3.7 Mpc).
{\tt Nuc} is the circle 
centered on the nucleus, {\tt diff-nuc} is the region surrounding this, 
{\tt diff-1} is the next region to the NW, and {\tt diff-2} is the largest region.
\label{fig:image}}
\end{figure}

The{\it  Chandra\/} data were taken on 27 - 28th January 2000 with the ACIS
camera in the faint mode for a total of 49 ks, with the nucleus
focussed on the S3 chip.  The filtered file produced by standard
processing was further cleaned by running the software tool {\tt
acisscreen} and gain corrected using {\tt acisgaincorr}.  These
cleaned data from the chip S3 were screened to reject periods of high
background (defined as times at which the total count rate of events
labelled as valid was greater than 20 counts per second), and the
resulting image of NGC~4945 with exposure of 38 ks is shown in
Fig.~\ref{fig:image}.  The image clearly shows a point-like emission
coincident with the megamaser source, and diffuse ``plume'' extending
roughly in the NW direction from the nucleus, which is perpendicular
to the plane defined by the megamaser emission of the galaxy, which in
turn is closely aligned with the plane of the host galaxy.  We perform
spectral analysis of various regions of that image separately as
described below.

The nuclear spectrum (hereafter called {\tt nuc}) was extracted from a
circle of radius 4 pixels (2 arcsec) centered on the brightest spot.
The nearby diffuse emission was taken from a box surrounding this
region, excluding a circle of radius 4.5 pixels centered on the
nucleus (hereafter called {\tt diff-nuc}). Further diffuse emission spectra
were taken from two regions shown overlaid on Fig.~\ref{fig:image},
hereafter called {\tt diff-1} and {\tt diff-2}.  
A background spectrum was taken
from a nearby, source free region.  The response and auxiliary files
were created for the nuclear region using {\tt acismakermf}, {\tt
acisarfprep} and {\tt mkarf}, and these files were used for all
spectra.  For the subsequent analysis, we grouped all extracted
spectra such that there were at least 20 total counts per new bin.

\subsection{RXTE PCA and HEXTE}

The simultaneous RXTE data were extracted using the rex script, 
with the Epoch 4 faint source background models. This resulted in a 
total of 60 ks of PCA data from layer 1, detectors 0 and 2.  As in 
the previous RXTE observations of this object, the source counting 
rate is rather modest, with source counts being less than 10 per cent 
of background.  

\begin{figure} 
\plotone{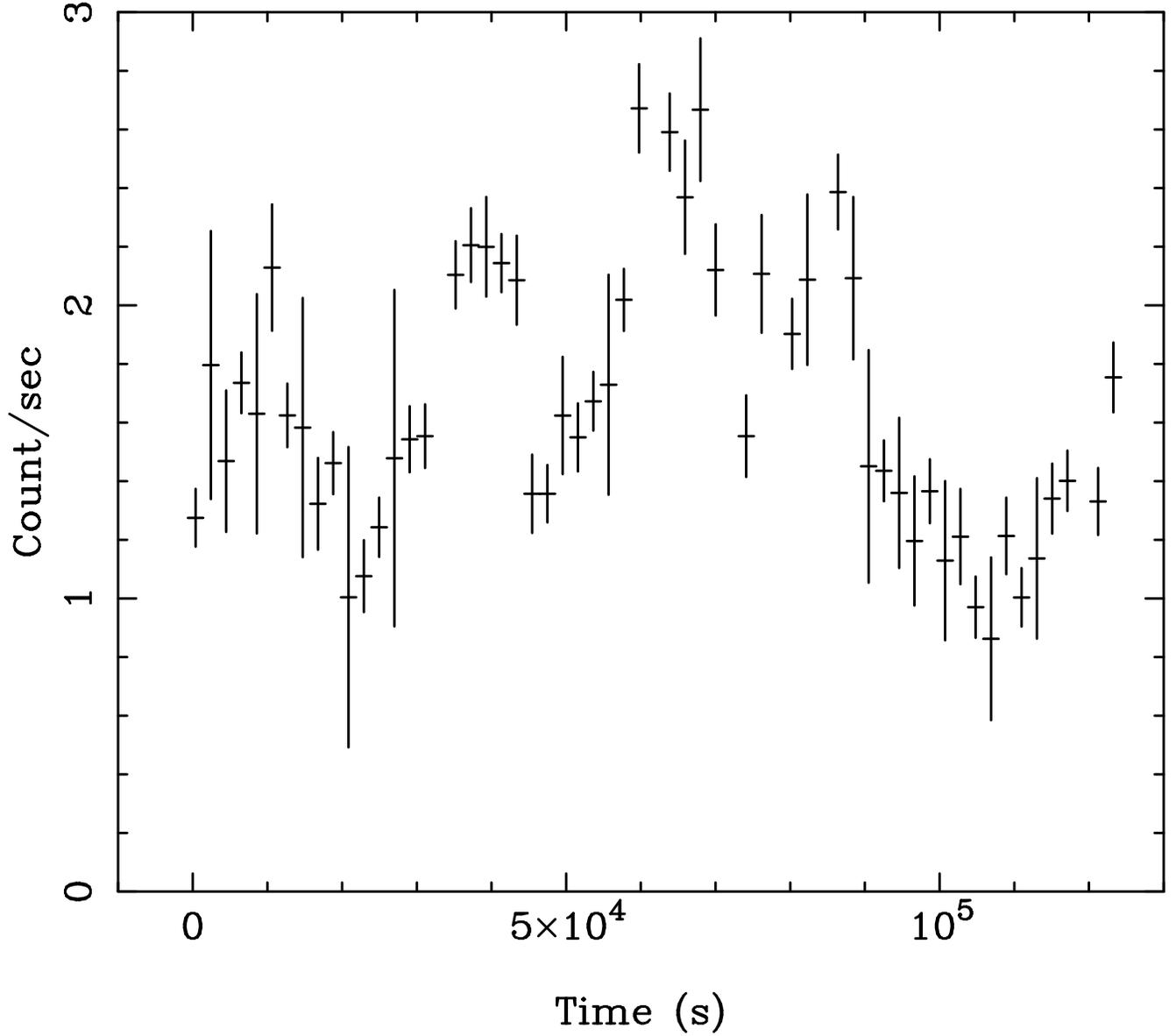}
\caption{PCA 8--30 keV lightcurve (channels 19-69), binned on 4096
second intervals. Continual variability is seen, showing that the optically
thick absorber cannot have a large scale height as electron scattering
would then smear out the rapid changes.
\label{fig:pcalc}}
\end{figure}

We know from previous observations that the nucleus of NGC~4945 is 
a relatively hard X-ray source, where the primary, nuclear emission 
can be well described as a heavily absorbed power law.  Since we are 
mainly interested in the RXTE PCA data regarding the nuclear
component, we present the PCA lightcurve in the 8--30 keV
band (channels 19 - 69) in Fig.~\ref{fig:pcalc}, with contiguous orbits
giving even sampling of the lightcurves on timescales of a few
thousand seconds, spanning a total of $\sim 1.5$ days. 
This extends the variability seen in the previous monitoring campaign,
which had single orbit snapshots once per day for $\sim 1.5$ months
(Madejski et al.\ 2000).  Plainly there is considerable variability 
power in this object on timescales shorter than 1 day.  

\begin{figure} 
\plotone{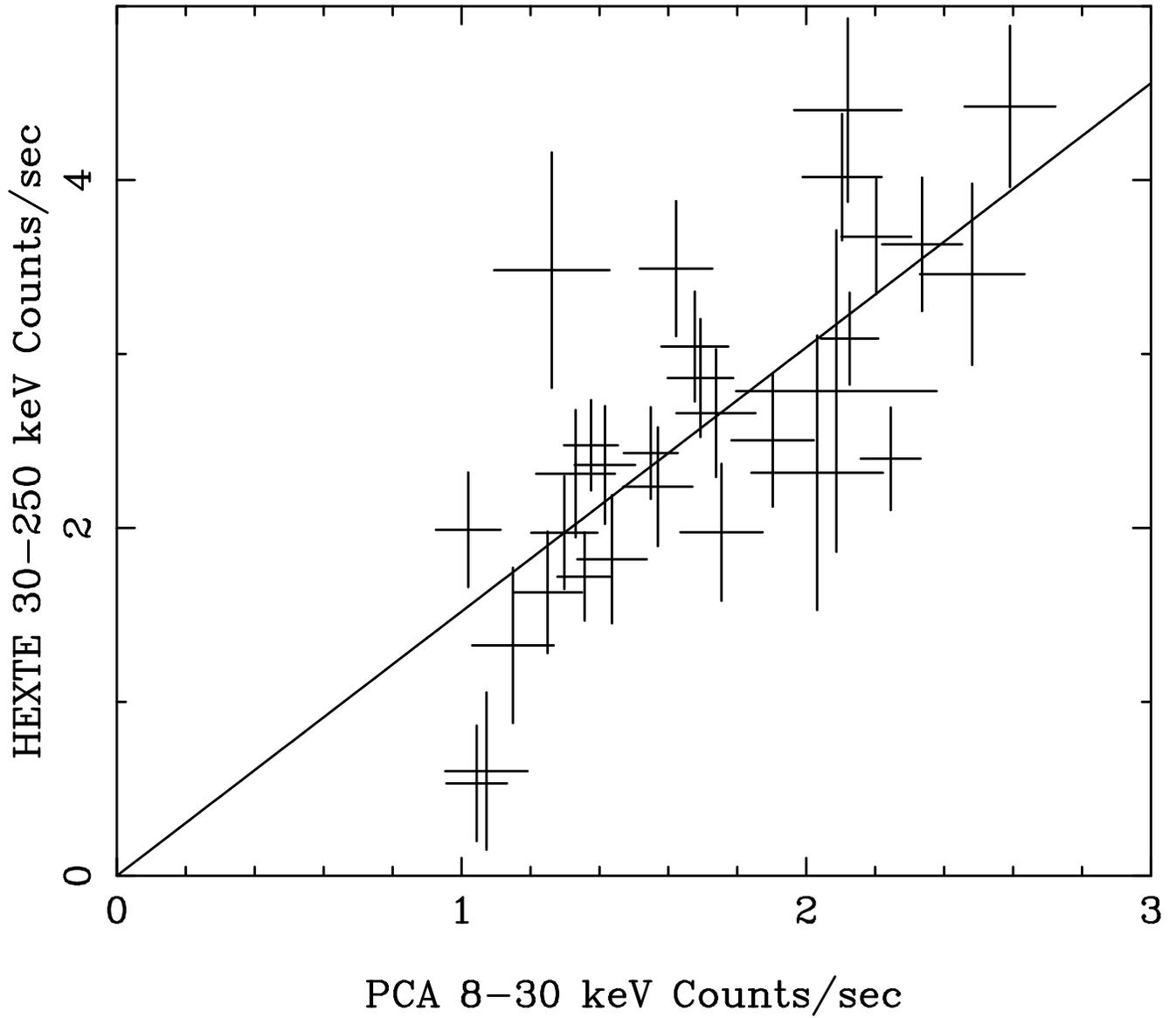}
\caption{RXTE PCA 8 -- 30 keV (channels 19 -- 69) lightcurve
versus the HEXTE 30 -- 250 keV (channels 15 -- 60) lightcurve, both binned
on 4096 second intervals. The higher energy variability is completely
consistent with the variability seen at lower energies.
\label{fig:pcahexte}}
\end{figure}

The HEXTE data from clusters 0 and 1 
were also extracted with the rex script, and here the
background is even more dominant.  Nonetheless, the variability seen in
the PCA and HEXTE are consistent with each other (Figure~\ref{fig:pcahexte}).
Since the method of background estimation is very 
different in the PCA (blank field predictions) and HEXTE (offset pointings)
then this shows that 
the majority of the variability seen is indeed connected to the
source rather than to background uncertainties. We use the HEXTE data
from 20--100 keV, and allow for a normalisation offset between
this and the PCA.

\section{Nuclear Spectrum}
\label{sec:nuclear}

\subsection{Chandra}

\begin{figure} 
\plotone{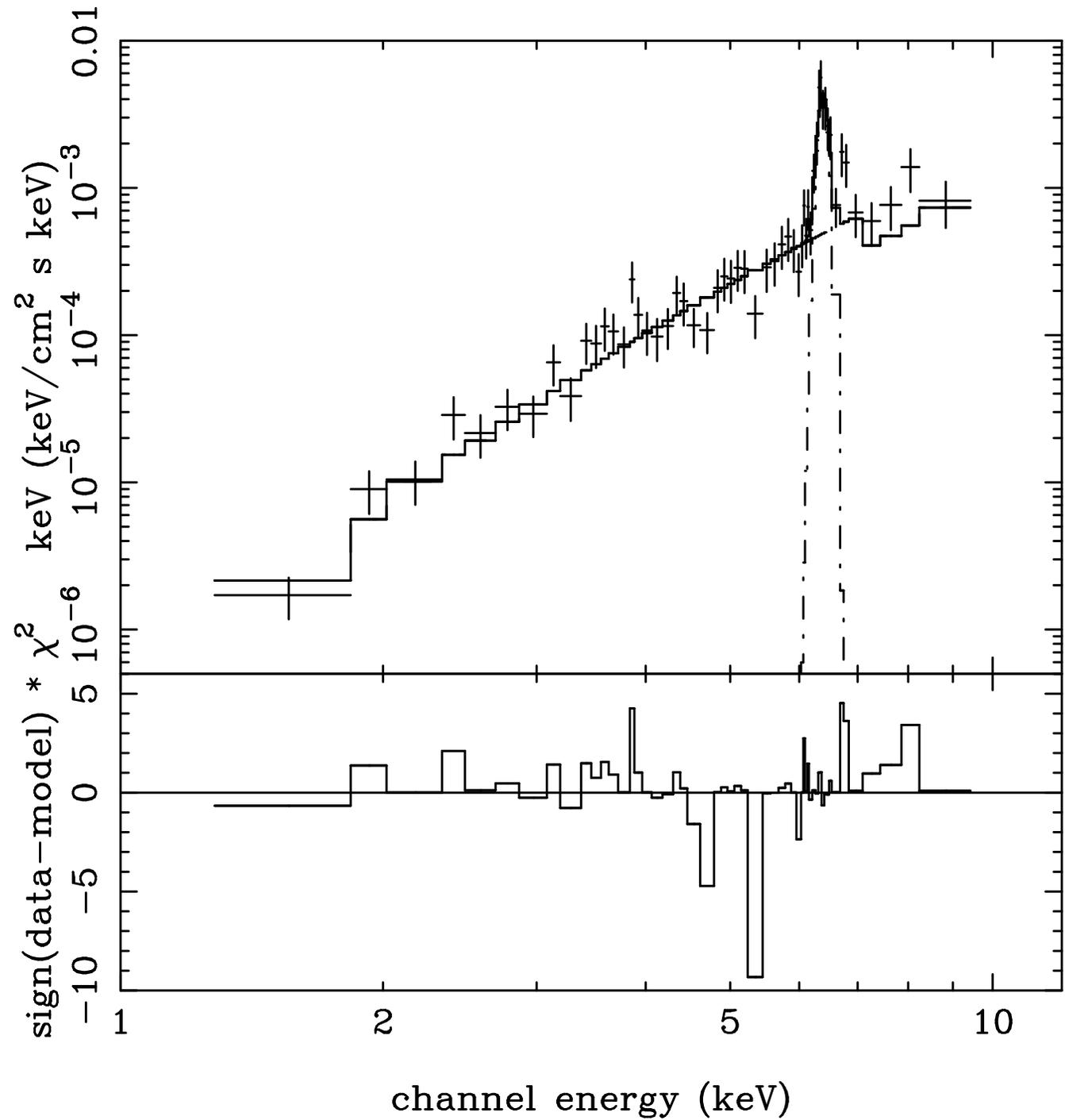}
\caption{The upper panel shows the 
Chandra ACIS spectrum from the nucleus, 
deconvolved with the reflection model. Residuals
to the fit are shown in the lower panel
\label{fig:nuclear}}
\end{figure}

The superb imaging capabilities of {\it Chandra\/} allow an extraction
of the spectrum from the nuclear source alone.  This is shown in
Fig~\ref{fig:nuclear} and is clearly dominated by iron K$\alpha$ line
emission, but also includes a hard broad-band continuum.  We fit this
with an absorbed power law and iron line and find that the nuclear
continuum is indeed extremely hard, with $\Gamma=-0.9$, and that the
(narrow, $\sigma$ fixed at 10 eV) line at 6.4 keV has a large
equivalent width of $1.3$ keV ($\chi^2_\nu=66.0/51$). % nuc_po_ga.xcm

The fit can be significantly improved if the line is broad, or if it
consists of a number of components.  Allowing the line to be broad we
obtain $\chi^2_\nu = 47.9/50$ with $\sigma=0.09^{+0.09}_{-0.03}$ keV,
and the EW increases to 2 keV with intensity $2.0\pm 0.4 \times
10^{-5}$ photons s$^{-1}$.  Alternatively, adding a second, narrow
line at 6.5 keV (fixed energy) gives $\chi^2=48.3/50$ and EW of 570 eV
and 270 eV for the 6.4 and 6.5 keV components, respectively. Repeating
the fits to ungrouped data, using C-statistics, gives (for 1043 PHA
bins): C-stat=929 for a narrow Gaussian at 6.38 keV, C-stat=906 for a
broad Gaussian and C-stat=911 for two narrow Gaussians. There is thus
a preference for a broad line, but it is rather marginal.

Clearly, the very hard power law component cannot be the nuclear
continuum observed directly.  Instead, in an attempt to construct a more
physical model of the source, we interpret the spectrum of
the compact region in the {\it Chandra\/} data to be solely due to Compton 
reflection of an unseen primary continuum.  
With this physical model
we obtain $\chi^2_\nu=65.2/53$   % nuc_felipl.xcm
for reflection of an intrinsic (unseen) power law of
$\Gamma=1.35^{+0.26}_{-0.22}$ by neutral, 
solar abundance material (assumed inclination of 60$^\circ$). 
The line emission in this model is calculated 
self-consistently with the
reflected continuum (\.{Z}ycki, Done \& Smith 1999),  so its equivalent
width is fixed with respect to the reflected continuum for a given 
illuminating power law and reflector ionization state and inclination
($\sim 1.4$ keV for the parameters used here).  The model
also includes the Compton downscattered shoulder on the iron K line,
so the line is intrinsically broad. However, the
%but this broadening is not sufficient to fit the observed line. The
fit can be significantly improved ($\chi^2_\nu = 56.3/52$)
by allowing the spectral
features to be further broadened, corresponding to a 
radius of $\sim 10^3\, {\rm R_g}$ if this is from Keplerian motion.

%The inferred intrinsic spectral index is very flat, but this is 
%consistent with the softer index of $\Gamma=1.85$ dervied for the
%overall spectrum as there is a 
%small contribution in the 8-10 keV band from the highly absorbed 
%nucleus (see next section and Fig.~\ref{fig:totnuc}).

The deconvolved spectrum with the unsmeared reflection model is
shown in Fig.~\ref{fig:nuclear}.  There is a marginally significant
residual ($\Delta\chi^2=5$ for 2 additional degrees of freedom) for a
narrow line at energy $6.73\pm 0.08$ which could indicate the
presence of reflection from more highly ionised material.

\subsection{Broad-band nuclear spectrum}
\label{sec:totnuc}

\begin{figure} 
\plotone{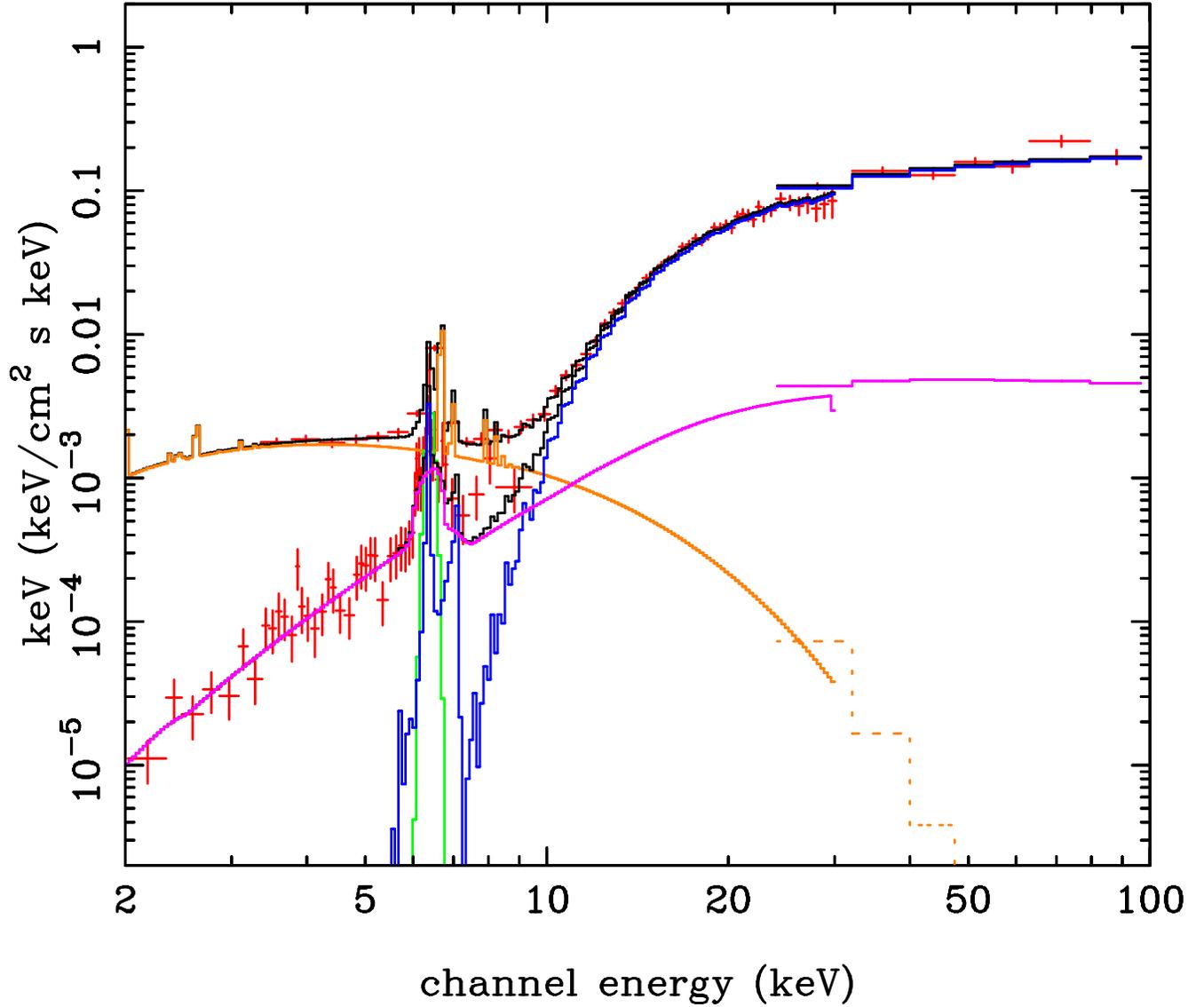}
\caption{The total nuclear spectrum as seen by Chandra, PCA and HEXTE.
The spectral data are unfolded from the instrument response using 
a model consisting of a heavily absorbed power law (blue)
which dominates above 15 keV in the PCA and HEXTE data.
Some fraction of the intrinsic (unabsorbed) spectrum is reflected from
cold material (magenta) and escapes along a path which does not intercept the 
extreme obscuring material, dominating the {\it Chandra} emission.
However, the wide field of view of the RXTE instruments
means that the PCA spectrum 
includes both diffuse emission and off-axis sources.
This additional X-ray flux is modelled by 
a hot plasma component which dominates the PCA data below 
$\sim 8$ keV (orange), together with an emission line at 6.4 keV (green).
\label{fig:totnuc} }
\end{figure}

Fig.~\ref{fig:totnuc} shows the broad band nuclear spectrum derived
from fitting the nuclear spectrum from {\it Chandra\/}, together with
the PCA and HEXTE data. {\it Chandra\/} can spatially resolve the
nuclear emission in the 1--10 keV range, but both the PCA and HEXTE
data cover a large field of view ($\sim 1 \times 1$ degree), so
include a contribution from off--nuclear point sources and host galaxy
diffuse emission as well as the nucleus itself. We fit the spectra
from the 3 instruments
simultaneously, but include a contribution from extended emission in
the PCA and HEXTE spectra which is set to zero in the {\it Chandra\/}
data. This extended emission can be well fit by a hot plasma
(modelled here using a solar abundance {\sc mekal} code), and we also 
include an additional emission line at 6.4 keV.
The nuclear emission is modelled by a heavily absorbed power law
and its weakly absorbed reflection ({\sc pexrav} plus a gaussian
line), and results are detailed in Table~\ref{tab:nuc_all}. 
The overall shape of the spectrum is very similar to that seen in
previous observations, and a direct
comparison of the PCA spectrum with that of Madejski et al. (2000) shows
no evidence for any changes in spectral shape, but the absorbed 
power law emission from the nucleus is a factor 1.8$\times$
brighter in the observations reported here.

The PCA spectrum contains much more line emission than seen in
Chandra.  While much of this is consistent with moderately ionised
(6.5--6.7~keV) emission from the hot diffuse plasma, there is also
evidence for some additional line at 6.4 keV, impling a contribution
to the fluorescent emission from the extended region (see also
Guainazzi et al. 2000; Schurch et al. 2002 and Section 4 below).

\begin{figure} 
\plotone{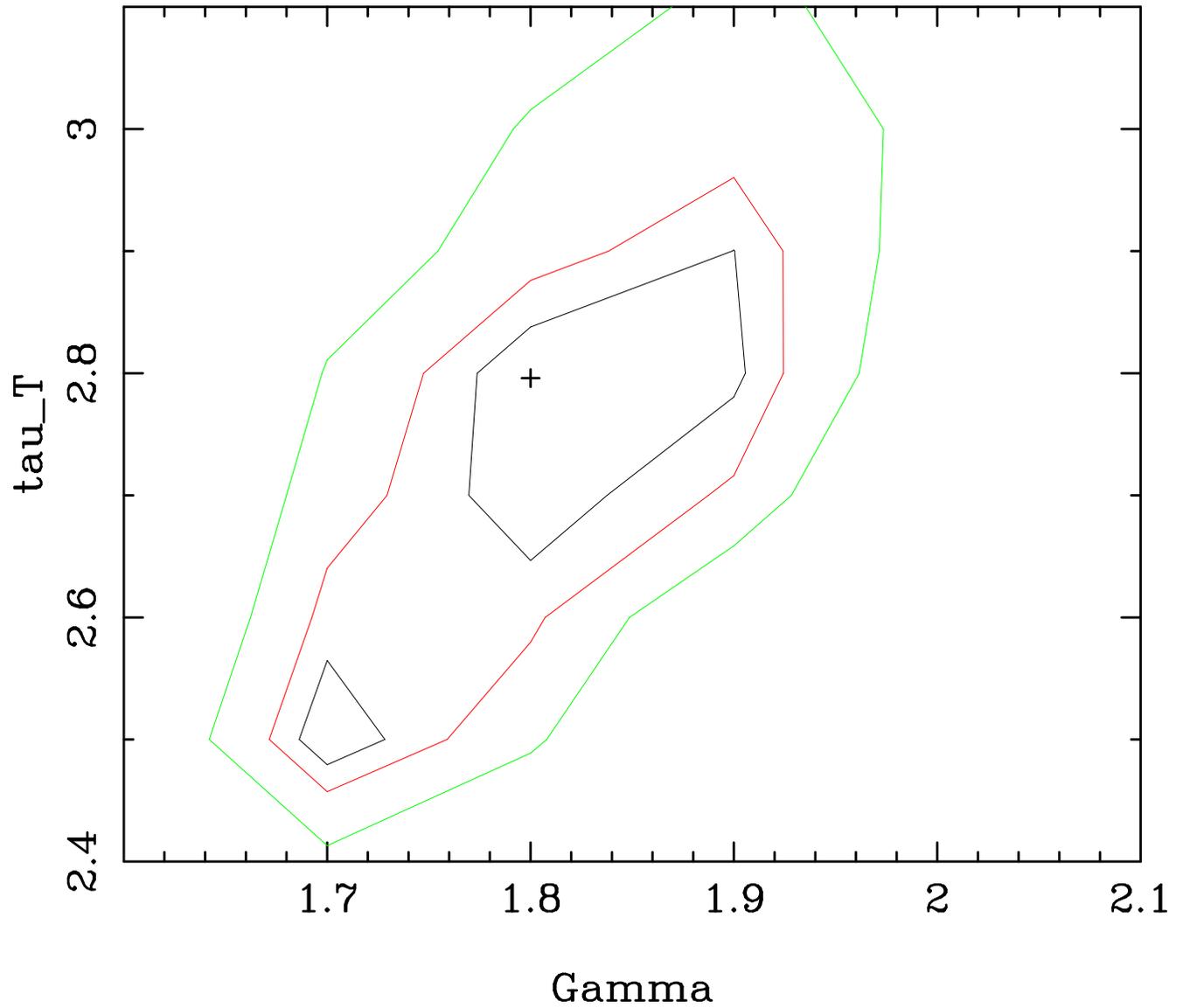}
\caption{Confidence contours on the intrinsic spectral index versus
the optical depth for the Monte Carlo absorption model shown in 
figure~\ref{fig:totnuc}. This relates to the standard column density 
through $\tau=1.21\times 6.65\times 10^{-25} N_H$
\label{fig:totcont}}
\end{figure}

The heavy absorption towards the nucleus implies that the obscuring
material is optically thick to electron scattering, with an optical
depth of a few.  This scattering changes the spectral shape from that
obtained by pure absorption, and we model this using the Monte Carlo
code of Krolik, Madau \& \.{Z}ycki (1994).  Motivated by previous and
current observations where the rapid variability implies a rather
geometrically thin (disk-like) absorbing structure, we assume that the
absorbing material subtends a rather small solid angle to the X-ray
source (Madejski et al.\ 2000). We assume this material forms a
symmetric torus, with half-angle of the absorber as seen from the
central source of $10^\circ$ i.e. a torus opening angle of $80^\circ$
which we view at $90^\circ$ (see Madejski et al.\ 2000).  We also
replace the separate reflection continuum/line model used above with
the self consistent reflected spectrum model.  This fit is detailed in
Table~\ref{tab:nuc_all}, and the resulting unfolded spectrum is
plotted in Fig.~\ref{fig:totnuc}. The overall spectral index is
$\Gamma\sim 1.8$ with 
confidence contours on the spectral index and optical depth
of the absorber are plotted in Fig.~\ref{fig:totcont}.

The optically thick absorption means that the true power law intensity
is suppressed by a factor $\sim exp(-\tau)$ even at 50-100 keV. The inferred
power law normalisation corrected for this large (but fairly
uncertain, see Fig.~\ref{fig:totcont}) factor
gives a 0.1-200 keV flux of $4.5\times
10^{-8}$ ergs cm$^{-2}$ s$^{-1}$. This implies an intrinsic X-ray flux of
$7\times 10^{43}$ ergs cm$^{-2}$ s$^{-2}$ in the 0.1--200 keV
bandpass, which is 50 per cent of the Eddington limit and 
similar to the observed FIR luminosity. Thus it is likely that the AGN
powers a substantial fraction of the observed FIR emission (Marconi et al. 2000).

\begin{deluxetable}{ccccccc} 
\label{tab:nuc_all}
\tablewidth{0pc} 
\tablecolumns{7} 
\tablecaption{
Results of modeling the nuclear spectrum from Chandra, PCA and HEXTE
simultaneous data}
\tablehead{ 
\colhead { $N_H$\tablenotemark{a} }  & 
$\Gamma$ & Norm & $N_H$\tablenotemark{b} & $f_{refl}$\tablenotemark{c}
& Fe K$\alpha$ intensity\tablenotemark{d} & $\chi^2$/dof 
}

\startdata

$425\pm 25$ & $1.65\pm 0.15$ & $0.03$ & $0.8\pm 0.8^{+1.2}_{-0.8}$ & $6\pm
2\times 10^{-2}$ & $2.0{+0.5}_{-0.3}\times 10^{-5}$ & $105.5/111$\tablenotemark{e} \\

\tablenotetext{a}{Absorption applied to the intrinsic nuclear (power
law) spectrum in units of $10^{22}$ cm$^{-2}$ }
\tablenotetext{b}{Absorption applied to all spectral components
in units of $10^{22}$ cm$^{-2}$ }
\tablenotetext{c}{Solid angle subtended by the reflector
(i.e. normalization relative to the direct power law normalization)
in units of $\Omega/2\pi$. Assumes a reflector inclination of
$60^\circ$.}
\tablenotetext{d}{6.4 keV iron line emission from the nucleus in 
photons cm$^{-2}$ s$^{-1}$}
\tablenotetext{e}{There is also a mekal plasma ($kT=5.3_{-1.1}^{+2.0}$~keV, 
2--10 keV unabsorbed flux of $5\times 10^{-12}$ ergs s$^{-1}$) 
and additional 6.4 keV narrow gaussian line with intensity $2.0\pm 0.9\times
10^{-5}$ photons cm$^{-2}$ s$^{-1}$ in the PCA and HEXTE data 
(normalizations set to zero in Chandra) so as
to account for the extended emission in their wide fields of view.}

\enddata 
\end{deluxetable}

\subsection{Origin of the nuclear 6.4 keV Fe K line}
\label{sec:fekline}

Clearly, one of the main questions is the origin of the Fe K line.  In
the model of the nuclear spectrum above the line can arise either in
the reflector or in the optically thick absorbing material.  This
distinction may be somewhat artificial as its possible to envisage a
geometry in which the absorber and reflector are the same structure,
e.g. where we are looking at an absorbing disk at an angle closer to
$80^\circ$ (the assumed opening angle) rather than $90^\circ$ so
reflected photons from the far side of the disk can be seen without
being absorbed. Any warp on the disk will also enhance the solid angle of
reflecting material which can be seen, and it is noteworthy that the
maser emission in this and other AGN indicate that the cool material
at 0.1--1~pc has a shallow warp. 

In our assumed geometry, where we view the absorbing 
disk/torus at $90^\circ$ then 
this material produces $\sim 25$\% of the total
line seen from the nucleus. The separate reflecting material produces
the rest of the line, and for solar abundances, the equivalent width
of the line to reflected continuum is about 1.3--1.6~keV (George \&
Fabian 1991; Matt, Perola \& Piro 1991).  This is not strongly
affected by increasing the abundances of all the heavy elements - this
increases the amount of line produced, but also increases the opacity
so fewer line photons escape (George \& Fabian 1991). However,
increasing the iron abundance relative to the other elements can give
a marked change in the line equivalent width (George \& Fabian 1991),
and fits to the {\it Chandra\/}/{\it RXTE\/} full nuclear continuum using
absorption/reflection models with iron alone at twice solar abundance
give a significantly worse fit than solar abundances. 
Iron overabundances are predicted in most chemical evolution 
models for AGN, and are often observed 
(Haman \& Ferland 1993). However, there is a delay of
$\sim 1-2$ Gyrs for the onset of the Fe producing 
Type 1a supernovae, so our observed solar abundances are
consistent with models for a young starburst 
(0.01 Gyr) in this object (Marconi et al. 2000).

Our reflection model assumes that the iron line is produced in
optically thick material. However, the line equivalent width can also
be affected by the column density of the material. The reflection
models assume that the material is optically thick, i.e. $N_H \ge
2\times 10^{24}$ cm$^{-2}$. At these columns then all the photons
above the iron edge which can produce the fluorescence line are
absorbed.  If the column is reduced below $\sim 10^{23}$
cm$^{-2}$ then this is no longer true. The material becomes thin to
the photoelectric absorption opacity at the iron edge and the line
decreases. But this also changes the reflected continuum - depending
on the geometry it can either look like a standard reflection spectrum
up to the point where the material becomes optically thin to
photoelectric opacity or it can look like straightforward absorption
by a column of $10^{23}$ cm$^{-2}$. Given that the observed {\em
continuum} in {\it Chandra\/} looks like optically thick reflection up to at
least 6-7 keV then the column density of the reflecting material {\em
must} be at least $10^{23}$ cm$^{-2}$.

\section{Diffuse emission}

\begin{figure} 
\plotone{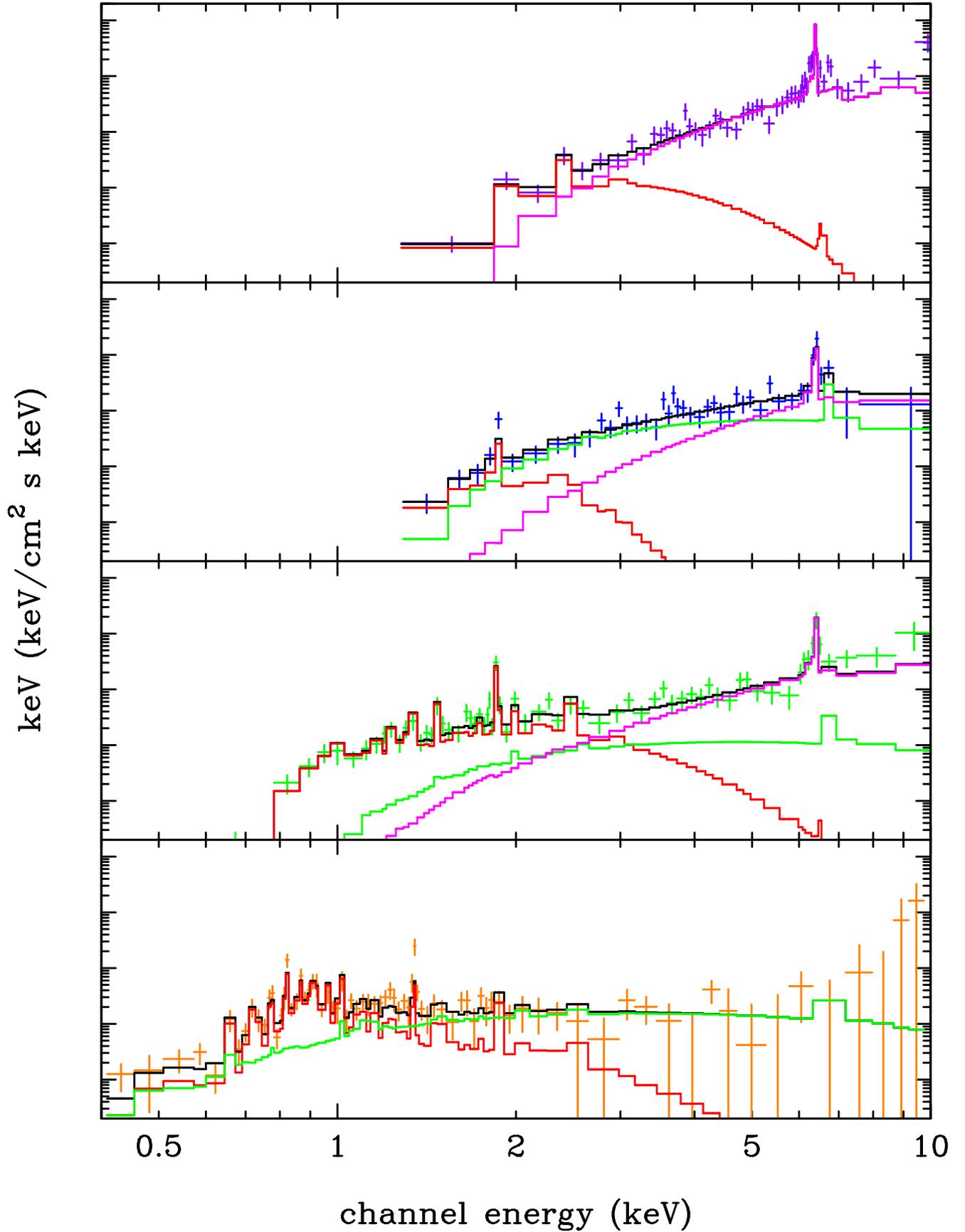}
\caption{Unfolded spectral data from the Chandra image shown in Figure 1.
The panels from top to bottom 
show the data from the nuclear region, {\tt diff-nuc}, {\tt diff-1}
and {\tt diff-2}, respectively. The model components are neutral
reflection (magenta), a low temperature {\sc mekal} plasma (red), 
and a high temperature {\sc mekal} plasma (green).
\label{fig:diffuse} }
\end{figure}

The {\it Chandra\/} spectra from the diffuse emission regions extracted from
regions located at progressively further distances from the nucleus
are shown in Fig.~\ref{fig:diffuse}.  The nuclear spectrum itself is
shown for comparison in the top panel. Plainly the diffuse spectra
nearby the nucleus ({\tt diff-nuc} and {\tt diff-1}) 
contain a substantial iron
K$\alpha$ line and show a hard continuum spectrum, which suggests that
they are also dominated by Compton reflection from optically thick
material at high energies. However, the increasing counts at low
energy, and increasing strength of line features (such as the $\sim
1.8 $ keV line, presumably due to He--like silicon, and the iron L
emission lines) show that there is
also an increasing fraction of the emission from hot/photoionized
plasma as distance from the nucleus increases. This hot plasma could
be either predominantly photoionised by the nucleus, or mechanically
heated by the starburst. In either case it will contain some free
electrons which scatter some fraction of the nuclear light, giving an
additional scattered power law component.

We assume the intrinsic primary emission is an isotropically emitted 
power law with fixed index $1.85$ and 
normalisation of $2.79$ photons cm$^{-2}$ s$^{-1}$ at 1 keV,
as derived from Section~\ref{sec:totnuc}. Some
fraction, $f_{refl}$ of this is reflected from cold material (again we
use the reflection code in which the self-consistent iron line
emission is included), while another fraction, $f_{scatt}$ is
scattered from hot electrons, forming a power law. 

We first assume that the line emission is from a mechanically heated
plasma (using the {\sc mekal} code), and these fits are detailed in
Table~\ref{tab:mekal}. We also include fits to 
the {\it Chandra\/} nuclear spectrum,
to show the limits on the scattered and hot gas emission on the
smallest scales, though here we truncate the data at 6.6~keV so as not
to include any transmitted flux. 
With this different energy range the
line broadening is much less significant, at $\Delta\chi^2=3$.

\begin{deluxetable}{ccccccc} 
\label{tab:mekal}
\tablewidth{0pc} 
\tablecolumns{7} 
\tablecaption{Results of modeling the diffuse emission data with the 
 {\sc mekal} model }
\tablehead{ 
\colhead {file}   & $N_H$ $\times 10^{22}$ cm$^{-2}$\tablenotemark{a} & kT (keV)
& mekal flux ergs s$^{-1}$\tablenotemark{b} &
$f_{scatt}$\tablenotemark{c} & 
$f_{refl}$ \tablenotemark{d} & $\chi^2$/dof }

\startdata

{\tt nuc} & 	   $7.1_{-3.5}^{2.9}$ & 0.7\tablenotemark{e} &
$2.4_{-2.4}^{+4.5} \times 10^{-12}$ & 
$0^{+0.8}\times 10^{-5}$ & $1.2_{-0.1}^{+0.2}\times 10^{-3}$ & 51.6/43\\
{\tt diff-nuc} & $3.8_{1.2}^{+1.3}$ & 0.7\tablenotemark{e} & $3.4_{-3.4}^{+9.6}\times 10^{-13}$ & 
$1.8_{-0.6}^{+0.7}\times 10^{-5}$ & $4.1\pm 1.1 \times 10^{-4}$ & 35.8/39\\
{\tt diff-1}	& $1.8\pm 0.2$ & $0.71\pm 0.11$ & $1.0_{-0.7}^{+0.4} \times 10^{-12}$ & 
$3.4_{-3.4}^{+4.6} \times 10^{-6}$ & $4.5_{-0.9}^{+0.5}\times 10^{-4}$ & 62.5/62\\
{\tt diff-2}	& $0.32^{+1.1}_{-0.7}$ & $0.68\pm 0.08$ & $1.0_{-0.3}^{+0.5}\times 10^{-13}$ & 
$5.1\pm 1.2\times 10^{-6}$ & $6.5_{-6.5}^{+400}\times 10^{-7}$ & 60.2/76\\

\tablenotetext{a}{Foreground absorption applied to all model
components}
\tablenotetext{b}{Bolometric, unabsorbed flux extrapolated over 0.01--100 keV}
\tablenotetext{c}{Normalization relative to that of the primary known from the broad  band modeling.}
\tablenotetext{d}{Normalization in units of $\Omega/(2\pi)$. Assumes 
inclination $60^\circ$.}
\tablenotetext{e}{Temperature fixed at 0.7 keV as the component is not
significantly detected.}

\enddata 
\end{deluxetable} 
 
The data show clear differences in the absorping column, decreasing as
a function of distance from the nucleus.  The ISM in our galaxy in
this direction has $N_H\sim 2\times 10^{21}$ cm$^{-2}$ while the
observed column is significantly higher than this in all fields except
{\tt diff-2}.  This is unsurprising as NGC~4945 is an edge on galaxy.
The {\it Chandra} nuclear spectrum is inferred to be $\sim 5\times$
more absorbed in these fits than from the reflection model
fitting in Section 3.1. This is due both to the
steeper illuminating spectrum and to ignoring the data above 6.6 keV
which include some component from the direct nuclear emission. 
Intruiguingly, CO observations imply a ring
of molecular gas with column of $\sim 7\times 10^{22}$ cm$^{-2}$ in
the nuclear direction (Whiteoak et al.\ 1990), and photoelectric
absorption is known to follow the molecular gas column in Seyfert 2's
(e.g. Wilson et al.\ 1998).

The data also indicate that the different spectral components behave
differently with distance from the nucleus. The neutral reflected
fraction is highest in the nucleus, then is smaller in {\tt diff-nuc}
and {\tt diff-1}, and much smaller (not significantly detected) in
{\tt diff-2}.  Conversely, the medium energy emission component
modelled here as scattered nuclear flux is not significantly detected
in the nucleus, is largest in {\tt diff-nuc} and {\tt diff-1}, while
it is smaller but still significant in {\tt diff-2}.  The low energy,
warm gas component is not significantly detected in {\tt nuc} or {\tt
diff-nuc}, but has a total luminosity of $\sim 10^{39}$ and $\sim
10^{38}$ ergs s$^{-1}$ in {\tt diff-1} and {\tt diff-2}, respectively.
The fits are acceptable in all cases, although line-like residuals
remain in {\tt diff-nuc} and {\tt diff-1}.

The self-consistency of this model can be checked by estimating the
density of the mekal plasma diffuse emission, and then using this to
calculate how important photo-ionisation should be.  The luminosity of
a mekal plasma of density $n$ in volume $V$ is $\Lambda n^2 V$, where
$\Lambda\sim 3\times 10^{-23}$ ergs cm$^{-3}$ s$^{-1}$ for a
temperature of $\sim 1$~keV. The three regions, {\tt diff-nuc}, {\tt
diff-1} and {\tt diff-2}, have volume of $\sim 5\times 10^{61}$,
$7\times 10^{61}$ and $10^{63}$ cm$^3$, respectively (assuming axial
symmetry), so the hot plasma densities are $\sim 1.7, 0.5$ and $0.07$
cm$^{-3}$ assuming it smoothly fills the volume.  In our model this
hot gas also scatters the direct nuclear flux. The scattered fraction
$f_{scatt}=\Omega/(4\pi) \tau =\Omega/(4\pi) n \sigma_T \Delta r $,
where the solid angle $\Omega/4\pi \sim 1, 0.27$ and $0.16$ for {\tt
diff-nuc}, {\tt diff-1} and {\tt diff-2}, and the path length $\Delta
r \sim 30, 86$ and $350$ pc. The scattered fraction predicted by the
hot plasma is $\sim 10^{-4}, 2.3\times 10^{-5}$ and $8\times
10^{-6}$. These are significantly bigger than the scattered fractions
derived from the fits, except for {\tt diff-2}. The fit result
scattered fractions predict densities of $\sim 0.3, 0.05$ and $0.06$
cm$^{-3}$, respectively, for plasma smoothly filling the volume.

The distance $r$ from the nucleus in each case is $\sim 50, 93$ and
$230$ pc, so the ionisation parameter, $\xi=L/(nr^2)$, is
approximately constant at $\sim 2000$ in each region. This high value
indicates that the model of collisionally heated hot gas filling the
volume is {\em not} self-consistent. Either the gas is strongly
clumped, so that its density is higher by a factor of $>10$, or
photoionisation will dominate over collisional equilibrium.  A key
problem with having photo-ionisation dominate is that there are {\em
no} signatures of the AGN illuminating the extended emission in any
other waveband. There are {\em no} standard AGN high excitation narrow
lines in the optical (e.g. [OIII] Moorwood et
al. 1996), near- or mid-IR (Genzel et al. 1998; Marconi et al. 2000:
Spoon et al. 2000), and the [FeII]/Br$\gamma$ line ratios are indicative of
shock heating rather than photoionisation (Reunanen, Kotilainen \&
Prieto 2002).  The observed extended emission line 
regions {\em must} be shielded
from the nuclear photoionising flux (5~eV--1~keV: Marconi et al. 2000),
but the extended 6.4 keV line emission shows that it {\em must} be
illuminated by photons $\ge 7$ keV. Hence the nucleus must be absorbed by
columns of $\ge 10^{22}$ cm$^{-2}$ in {\em all} directions.

Photo-ionisation by such a hard (absorbed) X-ray spectrum would lead to 
fluorescence lines from mostly neutral material rather than ionized
line emission. This is observed in
the extended 6.4 keV Fe line emission (significantly detected in
{\tt diff-nuc} and {\tt diff-1}), but this cannot explain the lower
energy line emission (e.g. Si at 1.8 keV or the iron L complex in {\tt
diff-1} and {\tt diff-2}). Thus the low energy lines must be from
collisionally ionised material, so the mekal parameters derived above
indicate that the material {\em must} be clumpy.

The requirement that the nuclear spectrum be absorbed means that the
model of a power law for the scattered flux is also not
consistent. Replacing this by an {\em absorbed} power law gives
stringent constraints in {\tt diff-2}. Absorption of $N_H\sim 1.4\times
10^{21}$ on the scattered flux increases $\chi^2$ by 2.7, while
columns of $10^{22}$ and $10^{23}$ cm$^{-2}$ increase it by 8.5 and
12.5, respectively. Thus it seems most likely that the 3--5 keV
continuum is {\em not} from scattered nuclear flux but rather is from
hot, clumped gas.

Models of starburst galaxies indicate that the material is indeed
strongly clumped (Suchkov et al. 1996; Strickland \& Stevens 2000),
such that multiphase and multitemperature gas exists at all radii.  We
replace the scattered component with a second, higher temperature
mekal plasma. This gives similar $\chi^2$ fits to all the spectra, but
requires temperatures of 4--7 keV in addition to the lower
temperature gas at 0.5--0.7 keV. Such hot gas is difficult to produce
(Suchkov et al. 1994; Strickland \& Stevens 2000), but this component
is observed in pure starburst galaxies (Pietsch et al. 2001), and a
similar model with multitemperature hot components was used by Schurch
et al. (2002) to fit the diffuse emission in both {\it Chandra\/} and 
{\it XMM\/}.
An alternative explanation could be that it is from an unresolved
population of X-ray binaries (Persic \& Rephaeli 2002). 

To summarize, {\em all} the off nuclear spectra require that the gas
is multiphase. The most likely interpretation is that they all have
multitemperature clumps of hot gas, while 
{\tt diff-nuc} and {\tt diff-1} also strongly require
the presence of cool, optically thick clumps.

\section{The Overall Geometry}

The {\it Chandra\/} image can be superimposed on previous images of this
galaxy at other wavelengths. Our {\tt diff-nuc} spectrum corresponds to the
100-200 pc edge-on starburst ring traced by molecular gas
(Br~$\gamma$: Moorwood et al.\ 1996, Pa~$\alpha$ and H$_2$: Marconi et
al.\ 2000 and CO: Curran et al.\ 2001). This ring has its major axis
along the major axis of the host galaxy (NE-SW direction), which also
matches the position angle of the central maser disk (Greenhill et
al.\ 1997). 

The {\tt diff-1} spectrum extends into the region of molecular gas where the
emission is dominated by H$_2$ rather than by Pa$\alpha$ (Marconi et
al.\ 2000). The Pa$\alpha$ line traces mainly starburst activity,
while the H$_2$ emission probably reflects shock heating on the edges
of the superwind cone (Moorwood et al.\ 1996).  The {\tt diff-2} spectrum
covers the extended narrow line emission seen as a cone in the low
excitation lines H$\alpha$ and [NII] (Moorwood et al.\ 1996).

The extreme absorption seen towards the nucleus in X-rays corresponds
to optical depths of $\sim 2.5$ to electron scattering. This material
has a rather small scale height as otherwise the scattered X-rays
would noticeably smear the hard X-ray variability (Madejski et al.
2000). A small, dense inner disk is also required to produce the
observed maser emission, so we identify the extreme absorber with the
masing disk. This material is distributed over a patch $\sim 0.7
\times 0.1$~pc in size (Greenhill et al. 1997), 
representing H$_2$ densities on the order of
$10^8$ to $10^{10}$~cm$^{-3}$ for fractional H$_2$O abundances of
$10^{-4}$ to $10^{-5}$ (Elitzurm 1992), so giving a potential column
of $N_H > 10^{26}$ cm$^{-2}$. Since this is considerably bigger than
the observed obscuration then we may not be in the maximally absorbed,
completely edge-on, line  of sight. 

In addition to the dense absorbing disk, the central engine must be
completely embedded in obscuring material within $\sim 25$~pc so that
UV and soft X-ray fluxes do not escape (Marconi et al. 2000).
Although this additional absorption could be associated with material
from the starburst-related inflow along a bar (Ott et al. 2001), we
suggest that it may be a high latitude extension of the dense absorber
(i.e., the maser-emitting material is a thickening in the equatorial
plane).  As such the central parsec of NGC\,4945 could be more gas
rich than that for most other Seyfert galaxies.  In this sense,
NGC\,4945 may be similar to NGC\,3079, another active galaxy that
exhibits very large columns (Iyomoto etal. 2001), substantial nuclear
star formation (Cecil et al. 2001 and references therein), and a
disordered but otherwise disk-like distribution of H$_2$O masers
(Trotter et al. 1998). This absorber is probably seen in the column of
$N_H\sim 7\times 10^{22}$ cm$^{-2}$ inferred on the reflected nuclear
X-ray spectrum, corresponding to $A_v\sim 35$. This is also the
absorbing column seen to the far IR nuclear source (Krabbe et
al. 2001).

On much larger scales there is absorption associated with 
dusty nuclear starburst ring which forms a $\sim$ 100-200 pc 
torus around the nucleus (Marconi et al. 2000). This picture fits into the 
growing evidence for two distinct absorption structures in many AGN,
with a compact, extreme absorption region surrounded by an
extended dusty lower absorption region (e.g. Granato et al. 1997).

The energy from supernovae in the starburst ring produces hot
multiphase gas which emits in the soft X-ray range.  A population of
X-ray binaries may also contribute to the spectrum from this region,
which would remove the requirement for some of this gas to be as hot
as $\sim 6$ keV. However, flux at $\sim 3-4$ keV is also required in
the superwind ({\tt diff-2}) region, where it is hard to envisage anything
other than hot gas being present.  This intrinsic diffuse emission is
augmented by some scattering of the absorbed nuclear flux in cold,
optically thick material. These cool clumps most probably represent
starforming cores as these have density $n\sim 10^6$ cm$^{-3}$ and
size scales of 1~pc (Plume et al. 1997).

\section{Conclusions}

The unsurpassed X-ray imaging ability of the {\it Chandra\/} satellite allows
us to disentangle the AGN and starburst/superwind contributions in
NGC~4945. {\it Chandra\/} sees the nucleus only in reflection: 
simultaneous {\it RXTE\/}
data show the direct nuclear flux absorbed by an extreme column of
$\sim 4\times 10^{24}$ cm$^{-2}$.

The starburst/superwind gas is clearly multi-phase, with cool clumps
(seen in reflection and iron fluorescence line emission at 6.4 keV)
co-existing with hot gas. The hot gas is itself clumped rather than
being uniform, as otherwise it would be strongly photo-ionised in
conflict with the observed spectrum from the furthest region
({\tt diff-2}). We show that the self-consistent scattered emission from the
hot gas is probably unimportant compared to its diffuse
emission. 

The extreme absorption of $N_H>10^{24}$ cm$^{-2}$
seen towards the nucleus cannot completely
cover the source in all direction as the 
extended 6.4 keV fluorescence line emission
clearly shows that the hard X-rays (7-10 keV) from the AGN illuminate the
starburst ring. However, the 
lack of optical/IR high excitation lines from this
region equally clearly shows that the UV/soft X-rays from the AGN do
not illuminate this material. Either the AGN does not produce UV/soft
X-rays (which seems highly unlikely) or they are absorbed in {\em all}
directions which requires columns of $N_H>10^{21-22}$ cm$^{-2}$
This implies that the nucleus is completely embedded in a
column of $\sim 10^{22-23}$ cm$^{-2}$, probably associated with
molecular gas driven into the nucleus by the starburst/superwind.

\section{Acknowledgements}

This work was supported by {\it Chandra\/} grant from NASA to Stanford 
University 
via the SAO award no. GO0-1038A and in part by Polish KBN grant 
2P03D01718.  PTZ and CD thank SLAC for the hospitality during visits
there.

\vfil\eject
\clearpage

\clearpage

\clearpage

\clearpage

\clearpage

\clearpage

\end{document}